\documentclass[twoside]{article}
\usepackage{qic}
\usepackage{epsfig}

\renewcommand{\Re}{\mathop{\mathrm{Re}}}
\renewcommand{\Im}{\mathop{\mathrm{Im}}}

\begin{document}
\setlength{\textheight}{8.0truein}    

\runninghead{SPIN SQUEEZING OF ONE-AXIS TWISTING MODEL...
$\ldots$} {Ji CG, $\ldots$}

\normalsize\textlineskip
\thispagestyle{empty}
\setcounter{page}{1}

\copyrightheading{13}{3\&4}{2013}{0266--0280}

\vspace*{0.88truein}

\alphfootnote

\fpage{1}

\centerline{\bf
{SPIN SQUEEZING OF ONE-AXIS TWISTING MODEL}} \vspace*{0.035truein}
\centerline{\bf {IN THE PRESENCE OF PHASE DEPHASING}}
\vspace*{0.37truein} \centerline{\footnotesize
CHEN-GANG JI $^1$, YONG-CHUN LIU$^1$\footnote{Present address: State
Key Lab for Mesoscopic Physics, School of Physics, Peking
University, Beijing 100871, China.}, and GUANG-RI JIN$^{1}$\footnote{Corresponding author: grjin@bjtu.edu.cn}}

\vspace*{0.015truein}
\centerline{\footnotesize\it $^1$Department of
Physics, Beijing Jiaotong University, Beijing 100044, China}

\baselineskip=10pt

\vspace*{0.225truein} \publisher{(July 11, 2012)}{(October 11, 2012)}

\vspace*{0.21truein}

\abstracts{
We present a detailed analysis of spin squeezing of
the one-axis twisting model with a many-body phase dephasing,
which is induced by external field
fluctuation in a two-mode Bose-Einstein condensates. Even in the
presence of the dephasing, our analytical results show that the
optimal initial state corresponds to a coherent spin state
$|\theta_{0}, \phi_0\rangle$ with the polar angle
$\theta_0=\pi/2$. If the dephasing rate $\gamma\ll S^{-1/3}$, where $S$ is total atomic spin, we
find that the smallest value of squeezing parameter (i.e., the
strongest squeezing) obeys the same scaling with the ideal
one-axis twisting case, namely $\xi^2\propto S^{-2/3}$.
While for a moderate dephasing, the achievable squeezing obeys the
power rule $S^{-2/5}$, which is slightly worse than the ideal
case. When the dephasing rate $\gamma>S^{1/2}$, we show that the
squeezing is weak and neglectable.}{}{}

\vspace*{10pt}

\keywords{Quantum noise, Bose-Einstein condensates, Phase dephasing, Spin squeezed states}
\vspace*{3pt}
\communicate{D Wineland \& K Moelmer}

\vspace*{1pt}\textlineskip    

\section{\label{Sec:intro}Introduction}


Spin squeezing of an ensemble of spin-$1/2$ particles have attracted considerable attention for decades because they are not only important in view point of fundamental physics but also have a lot of applications. For instance, a quantum interferometer felt with spin squeezed states (SSS) or
multi-particle entangled states (MES) in the input ports can
improve phase sensitivity beyond standard quantum limit
(SQL)~\cite{Caves,Yurke,Wineland}. Dynamical generation of the SSS~\cite{Kitagawa,Ma} and the
MES~\cite{Molmer,You03a,Pezze&Smerzi} has been proposed by using the `one-axis
twisting' (OAT) interaction, which leads to quantum correlation among individual spin in a collective spin system. In addition, the OAT scheme of spin squeezing can be transformed into the so-called two-axis twisting with a sequence of $\pi/2$ pulses~\cite{YCL&YL}.

Under governed by the OAT Hamiltonian, the spin system can evolve from
a coherent spin state (CSS) into a spin squeezed state, which
shows the reduced variance $V_{-}$ below standard quantum limit
(SQL)---$S/2$, where $S$ is total atomic spin. The degree of spin
squeezing $\xi^2$ ($=2V_{-}/S$) can reach the power rule
$S^{-2/3}$, which is the ideal OAT result~\cite{Kitagawa}.
Experimental realizations of the OAT model has been
proposed~\cite{Milburn,Cirac,Sorensen,Bigelow,Law,Jin07,Jin08,YC} and
demonstrated~\cite{Orzel,Greiner,Strabley,Jaksch,Esteve,Chuu,Jo,Gross,Riedel,Maussang}
in a two-mode Bose-Einstein condensate (BEC). Due to experimental
imperfections in coupling pulses, atom losses, technique and
quantum noises, etc., the achievable squeezing is worse than the
theoretical prediction~\cite{LiYun}. Recently, we investigated the dependence
of spin squeezing on the initial CSS $|\theta_0, \phi_0\rangle$. Our
results show that the scaling of $\xi^2$ depends sensitively upon the polar angle $\theta_0$; it becomes $\xi^2\propto
S^{-1/3}$~\cite{Jin09} even when $\theta_0$ is
slightly deviated from its optimal value $\pi/2$.

In this paper, we investigate the effect of phase dephasing on
spin squeezing of the OAT model. The dephasing process considered
here is induced by external field
fluctuation~\cite{Gardiner,Milburn08,Huelga,Kitagawa01,Rey,Boixo08,Khodorkovsky,ljy,Jin10},
which gives rise to an enhanced decay of the phase coherence
between two modes of the BEC~\cite{Artur}. We present the details
of Ref.~\cite{Jin09} to obtain analytical results of the strongest
squeezing $\xi^2_{\min}$. For the optimal initial CSS with the
polar angle $\theta_0=\pi/2$, we find that the dephasing effect
can be negligible as long as the dephasing rate $\gamma\ll
S^{-1/3}$. The ideal one-axis twisting is attainable with the best squeezing $\xi^2_{\min}\propto
S^{-2/3}$~\cite{Kitagawa}. For a moderate dephasing rate (i.e.,
$S^{-1/3}<\gamma<S^{1/2}$), our analytical result indicates that
the achievable squeezing scales as $S^{-2/5}$, which is slightly
worse than the ideal OAT case. As the dephasing rate increases up
to $S^{1/2}$, we find that the squeezing becomes very weak as
$\xi^2\sim 1$.

This paper is arranged as follows.
In Sec.~II, we consider quantum dynamics of the one-axis twisting model with a many-body phase dephasing, which describes quantitatively a two-mode BEC in the presence of the phase diffusion. The density-matrix elements are obtained by solving quantum master equation. In Sec.~III, we investigate the OAT-induced spin squeezing by calculating exact solutions of five relevant quantities, $\langle
\hat{S}_{z}\rangle$, $\langle\hat{S}_{+}\rangle$, $\langle\hat{S}_{z}^{2}\rangle$, $\langle\hat{S}_{+}^{2}\rangle$, and
$\langle\hat{S}_{+}(2\hat{S}_{z}+1)\rangle$. With these expectation values at hands, one can numerically calculate the squeezing parameter $\xi^2$. To get scaling behavior of $\xi^2$, in Sec.~IV, we consider short-time limit and large enough particle number. Analytical expression of the squeezing parameter is obtained, with which we analyze power rules of $\xi^2$ according to the role of the phase dephasing. Finally, we
conclude in Sec.~V with the main results of our work.

\section{The one-axis twisting Model}

We focus on a two-component Bose-Einstein condensates with the internal
states $|\uparrow\rangle$ and $|\downarrow\rangle$ that is confined in a
deep three-dimensional harmonic potential. Quantum dynamics of the total
system can be described by the Lindblad equation ($\hbar =1$):
\begin{equation}
\frac{d\hat{\rho}}{dt}=i[\hat{\rho},\hat{H}]+\Gamma
_{p}(2\hat{S}_{z}\hat{\rho}\hat{S}_{z}-\hat{S}_{z}^{2}\hat{\rho}-\hat{\rho}\hat{S}_{z}^{2}),
\label{MSE}
\end{equation}
where $\hat{H}=\chi\hat{S}_{z}^{2}$, known as the one-axis
twisting Hamiltonian~\cite{Kitagawa}, can be realized in a
two-mode BECs~\cite{Milburn,Cirac,Sorensen} with the interaction strength $\chi$
tunable via the Feshbach resonance technique~\cite{Gross} or the BEC spatial
splitting~\cite{Riedel}. Atomic spin operators $\hat{S}_{z}=(\hat{b}_{\uparrow}^{\dag}\hat{b}_{\uparrow}-\hat{b}_{\downarrow}^{\dag
}\hat{b}_{\downarrow})/2$ and $\hat{S}_{+}=(\hat{S}_{-})^{\dag
}=\hat{b}_{\uparrow}^{\dag}\hat{b}_{\downarrow}$ represent
atomic population imbalance and phase coherence between the two
bosonic modes. The second term in Eq.~(\ref{MSE}) simulates a
phase dephasing of the BEC due to magnetic-field
fluctuations~\cite{Gardiner,Milburn08,Huelga,Kitagawa01,Rey,Boixo08,Khodorkovsky,ljy,Jin10,Artur}.
Such a kind of many-body decoherence has also been studied in
cavity-QED system~\cite{Agarwal,Chen}.

Due to particle-number conservation, the OAT Hamiltonian and the
super-operator in the Eq.~(\ref{MSE}) commutes with total angular
momentum operator $\mathbf{\hat{S}}^{2}$. Consequently, the state
at any time $t$ can be expanded in terms of common eigenstates of
$\mathbf{\hat{S}}^{2}$ and $\hat{S}_{z}$, i.e., $\{|S, m\rangle\}$
with total angular momentum $S=N/2$ and $m=-S$, $-S+1$, $\cdots$,
$S$. Using the basis, the density-matrix operator $\hat{\rho}$
reads
\begin{equation}
\hat{\rho}=\sum_{m, n=-S}^{S}\rho_{m, n}\left\vert S, m\right\rangle
\left\langle S, n\right\vert,  \label{rho}
\end{equation}%
with the elements $\rho_{m,n}=\langle S, m|\hat{\rho}|S, n\rangle$ satisfying
\begin{equation}
\frac{d\rho_{m,n}}{dt}=\left[i\chi(n^{2}-m^{2})-\Gamma_{p}(n-m)^{2}\right]\rho _{m,n}. \label{MSE2}
\end{equation}%
Assume that the BEC system evolves from a coherent spin state
(CSS)~\cite{CSS}: $|\theta_{0}, \phi_{0}\rangle\equiv \exp
\{i\theta_{0}[\hat{S}_{x}\sin(\phi_{0})-\hat{S}_{y}\cos(\phi_{0})]\}|S, S\rangle =\sum_{m}c_{m}|S, m\rangle$, with the
probability amplitudes
\begin{equation}
c_{m}={2S\choose S+m}^{1/2}\left(\cos\frac{\theta_{0}}{2}\right)^{S+m}
\left(\sin\frac{\theta_{0}}{2}\right)^{S-m} e^{i(S-m)\phi_{0}}.
\label{cm}
\end{equation}
Exact solutions of Eq.~(\ref{MSE2}) can be obtained with the density-matrix
elements
\begin{equation}
\rho_{m,n}(\tau)=\rho_{m,n}(0)e^{-i(m^{2}-n^{2})\tau}e^{-(m-n)^{2}\gamma\tau},  \label{rhodp}
\end{equation}
where $\tau=\chi t$ and $\gamma=\Gamma_{p}/\chi$. The first term on the right-hand side of Eq.~(\ref{rhodp}) $\rho_{m, n}(0)=c_{m}c_{n}^{\ast}$, with $c_{m}$ given by
Eq.~(\ref{cm}). The second term arises from time evolution of the density matrix under the OAT Hamiltonian $\chi \hat{S}_z^2$. The last one is the dephasing term due to magnetic-field fluctuation and has been obtained previously~\cite{Takeuchi,Genoni,Ferrini,Sinatra1}. Particularly, Takeuchi {\it et al.}~\cite{Takeuchi} considered a light-induced spin squeezing in an atomic gas and obtained almost the same result with ours. However, their result is based upon an approximated commutation relation of the Stokes operators of light. Here we present exact solution of the density-matrix elements, which describes quantitatively the OAT-induced spin squeezing in a two-mode BEC in the presence of the phase dephasing.

\section{Spin squeezing of the OAT model}

Starting from a CSS $|\theta_{0}, \phi_{0}\rangle$, unitary
evolution of the spin system under the OAT Hamiltonian leads to
spin squeezing and multipartite entanglement. Both of them can be
quantified by the variances of a spin component $\hat{S}_{\psi
}=\mathbf{\hat{S}}\cdot\mathbf{n}_{\psi}$ that is normal to
the mean-spin vector $\langle\mathbf{\hat{S}}\rangle\equiv (\langle
\hat{S}_{x}\rangle, \langle\hat{S}_{y}\rangle, \langle
\hat{S}_{z}\rangle)$, i.e., $\mathbf{n}_{\psi}\cdot\langle
\mathbf{\hat{S}}\rangle=0$. As usual, the expectation value of an operator
$\mathcal{\hat{O}}$ is defined by $\langle
\mathcal{\hat{O}}\rangle
=\mathrm{Tr}(\hat{\rho}\mathcal{\hat{O}})$. For a given state $\hat{\rho}$, the
mean spin and its direction $\mathbf{n}_{3}=\langle
\mathbf{\hat{S}}\rangle/|\langle \mathbf{\hat{S}}\rangle |$ can
be determined uniquely, which in turn gives $\mathbf{n}_{\psi}=\mathbf{n}_{1}\cos\psi +\mathbf{n}_{2}\sin \psi
$, with three orthogonal vectors
$\mathbf{n}_{1}$, $\mathbf{n}_{2}$, and
$\mathbf{n}_{3}$ (for details see Ref.~\cite{Jin09}). The increased and the reduced variances of the spin component
$\hat{S}_{\psi}$ are defined, respectively, as $V_{+}=\max_{\psi}(\Delta
\hat{S}_{\psi})^{2}$ and $V_{-}=\min_{\psi}(\Delta\hat{S}_{\psi})^{2}$, with
\begin{equation}
V_{\pm}=\frac{1}{2}\left(\mathcal{C}\pm
\sqrt{\mathcal{A}^{2}+\mathcal{B}^{2}}\right) ,  \label{V+-}
\end{equation}
where $\mathcal{A}$, $\mathcal{B}$, and $\mathcal{C}$ depend only
on five expectation values (see Appendix A): $\langle
\hat{S}_{z}\rangle$, $\langle\hat{S}_{+}\rangle$, $\langle\hat{S}_{z}^{2}\rangle$, $\langle\hat{S}_{+}^{2}\rangle$, and
$\langle\hat{S}_{+}(2\hat{S}_{z}+1)\rangle$. According to
Kitagawa and Ueda~\cite{Kitagawa}, a spin state is squeezed only
if the variance $V_{-}$ is smaller than the SQL, $S/2$, namely the
squeezing parameter
\begin{equation}
\xi^{2}=\frac{2V_{-}}{S}<1.  \label{xi2}
\end{equation}
The spin squeezed state is
useful to improve frequency resolution in spectroscopy provided
that $\zeta_{\mathrm{S}}^{2}=2SV_{-}/|\langle
\mathbf{\hat{S}}\rangle|^{2}<1$~\cite{Wineland}, which provides a
sufficient criterion for the degree of multipartite
entanglement~\cite{Sorensen}.
In addition, the squeezing parameter can be defined as $\zeta^2=2V_{-}/|\langle
\mathbf{\hat{S}}\rangle|=(S/|\langle
\mathbf{\hat{S}}\rangle|)\xi^2$~\cite{Satio}. The three definitions are slightly different in magnitude, $\xi^2\leq\zeta^2\leq\zeta_{\mathrm{S}}^{2}$ due to $|\langle
\mathbf{\hat{S}}\rangle|\leq S$. For large enough particle number $N$ ($>10^2$), the minimum values of them obey almost the same power rule~\cite{Takeuchi}, so we only focus on the squeezing parameter $\xi^2$.

\begin{figure} [hpbt]
\centerline{\epsfig{file=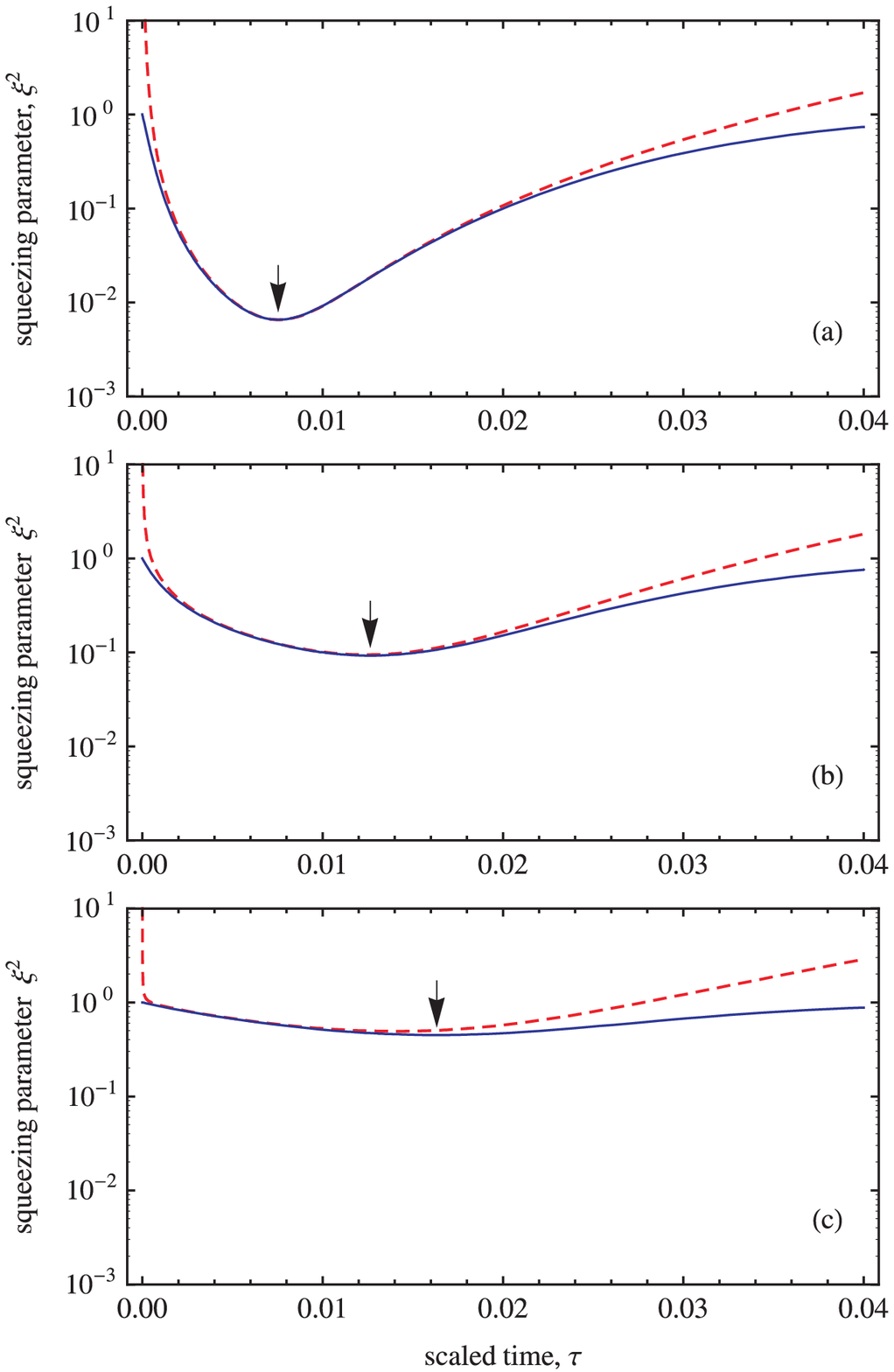, width=9.5cm}} 
\vspace*{13pt} \fcaption{\label{fig1}(color online) The squeezing parameter $\xi^{2}$ as a
function of scaled time $\tau$ ($=\chi t$) for various
dephasing rates $\gamma=\Gamma_p/\chi=0$ (a), $1$ (b), and $10$
(c). Solid blue lines are given by exact numerical simulations and
red dashed lines are predicated by Eq.~(\ref{xi2ana}). Other
parameters are $S=N/2=10^{3}$, $\theta_{0}=\pi/2$, and $\phi_{0}=0$. The arrows indicate the location of maximal-squeezing times $\tau_{\min}=7.536\times 10^{-3}$ (a), $1.264
\times 10^{-2}$ (b), and $1.632\times
10^{-2}$ (c), at which the strongest squeezing occurs with $\xi_{\min}^{2}=6.56\times 10^{-3}$ (a), $9.278\times 10^{-2}$ (b), and $0.4504$ (c).}
\end{figure}

Based upon
Eq.(\ref{rhodp}), we now calculate exact solutions of the mean spin
and the variances $V_{\pm}$. After some straightforward calculations, we can obtain (see Append.~A, or
Ref.\cite{Jin09})
\begin{equation}
\langle\hat{S}_{z}\rangle=S\cos(\theta_{0}), \hskip8pt \langle
\hat{S}_{+}\rangle=S\sin(\theta_{0})e^{i\phi_{0}}e^{-\gamma\tau}[R(\tau)]^{2S-1},
\label{JzJ+}
\end{equation}
where the population imbalance $\langle\hat{S}_{z}\rangle$ is time-independent, and
\begin{equation}
R(\tau)=\cos(\tau)+i\cos(\theta_{0})\sin (\tau)=\sqrt{1-\sin^{2}(\theta_{0})\sin^{2}(\tau)}\cdot e^{i\tan
^{-1}[\cos(\theta_{0})\tan(\tau)]}.  \label{f(t)}
\end{equation}
From Eq.~(\ref{JzJ+}), we note that the phase dephasing
considered in Eq.~(\ref{MSE}) imposes an exponential decay term
$e^{-\gamma\tau}$ to the phase coherence $|\langle S_{+}\rangle|$, but maintains the population imbalance $\langle
\hat{S}_{z}\rangle$ intact. Moreover, it is easy to obtain
$\langle\hat{S}_{x}\rangle=\Re\langle\hat{S}_{+}\rangle =|\langle\hat{S}_{+}\rangle|\cos(\phi)$ and $\langle\hat{S}_{y}\rangle=\Im
\langle\hat{S}_{+}\rangle=|\langle\hat{S}_{+}\rangle|\sin(\phi)$, with the
argument of $\langle\hat{S}_{+}\rangle$:
\begin{equation}
\phi\equiv\arg\langle\hat{S}_{+}\rangle=\phi_{0}+(2S-1)\tan^{-1}[\cos
(\theta_{0})\tan(\tau)].  \label{phi}
\end{equation}%
Here, $\phi _{0}$ is the azimuth angle of the initial CSS. The
variances $V_{\pm }$ depend upon the coefficients $\mathcal{A}$,
$\mathcal{B}$, and $\mathcal{C}$ (see Append. A). In real
calculations of them, only $\cos(\phi)$ and $\sin(\phi)$ are required, which depends on $\langle\hat{S}_{+}\rangle$. In addition, we need to solve the following
expectation values:
\begin{equation}
\langle \hat{S}_{z}^{2}\rangle=\frac{S}{2}+S\left( S-\frac{1}{2}\right)
\cos ^{2}(\theta_{0}),  \label{Jz2exact}
\end{equation}
\begin{equation}
\langle \hat{S}_{+}^{2}\rangle=S\left( S-\frac{1}{2}\right) \sin
^{2}(\theta _{0})e^{2i\phi
_{0}}e^{-4\gamma \tau }[R(2\tau )]^{2S-2},  \label{S+2exact}
\end{equation}
and
\begin{eqnarray}
\langle \hat{S}_{+}(2\hat{S}_{z}+1)\rangle &=&2S\left(S-\frac{1}{2}\right)
\sin (\theta_{0})e^{i\phi
_{0}}e^{-\gamma\tau}[R(\tau )]^{2S-2} [i\sin\left(\tau\right)+\cos(\theta_{0})\cos\left(\tau \right)].  \label{S+2Jz+1exact}
\end{eqnarray}
Substituting the above results into the coefficients
$\mathcal{A}$, $\mathcal{B}$, and $\mathcal{C}$, one can obtain
the variances $V_{\pm}$ and also the squeezing
parameter $\xi^{2}$. In Fig.~\ref{fig1}, we plot exact numerical
results of $\xi^{2}$ (solid lines) for different values of the
dephasing rate $\gamma$. We find that the strongest squeezing occurs at a certain time $\tau_{\min}$ ($=\chi t_{\min}$), with its position indicated by the arrows of Fig.~\ref{fig1}.

\section{Scaling behaviors of the squeezing parameter}

In order to analyze scaling behaviors of $\xi^{2}$, we now
consider the short-time limit (i.e., $\tau=\chi t\ll 1$) and
large enough particle-number $(S=N/2\gg 1)$, so Eq.~(\ref{f(t)})
can be approximated as $R(\tau)\!\approx \! S\exp
(-\frac{1}{2}\tau^{2}\sin^{2}\theta_{0})e^{i\tau\cos\theta_{0}}$, which in turn yields
\begin{equation}
\langle\hat{S}_{+}\rangle \!\approx \!S\sin (\theta_{0})e^{i\phi}e^{-\beta},  \label{S+a}
\end{equation}
\begin{equation}
\langle\hat{S}_{+}^{2}\rangle \!\approx \!S\left(S-\frac{1}{2}\right)\sin^{2}(\theta_{0})e^{2i\phi}e^{-4\beta },  \label{S+2a}
\end{equation}
and
\begin{equation}
\langle \hat{S}_{+}(2\hat{S}_{z}+1)\rangle \!\approx 2S\left(S-\frac{1}{2}\right)\sin(\theta_{0})e^{i\phi}e^{-\beta}\left(
\cos\theta_{0}+i\tau\right) ,  \label{S+2Jz+1a}
\end{equation}
where the argument of $\langle\hat{S}_{+}\rangle$ now becomes $\phi\approx
\phi_{0}+2S\tau\cos(\theta_{0})$, and
\begin{equation}
\beta=S\tau^{2}\sin^{2}\theta_{0}+\gamma\tau .  \label{beta}
\end{equation}
Note that without the dephasing, the phase coherence reduces to $|\langle S_{+}\rangle|\approx\!S\sin(\theta_{0})e^{-(\tau/\tau_{d})^{2}}$, with
the coherent time $\tau_{d}=\chi t_{d}=1/(\sqrt{S}\sin\theta
_{0})$~\cite{Jin09}. Such a kind of exponential decay is
well known as the so-called phase diffusion of a two-mode BEC. For the BEC atoms, the nonlinearity $\chi\propto\frac{1}{2}(a_{\uparrow\uparrow}+a_{\downarrow\downarrow
}-2a_{\uparrow\downarrow})$ depends upon the intra- and the
inter-species atom-atom scattering lengthes. When the three
coupling constants are close to each other, the coherence time
$t_{d}$ increases dramatically due to $\chi\rightarrow
0$~\cite{Sinatra}. If we take the phase dephasing into account
(i.e., $\gamma\neq 0$), the phase diffusion process will speed
up, as demonstrated recently in Ref.~\cite{Artur}. In what's following,
we will investigate the role of the phase dephasing on the spin
squeezing.

Firstly, using Eq.~(\ref{S+2a}) and Eq.(\ref{S+2Jz+1a}), as well as the exact result of $\langle\hat{S}_{z}^{2}\rangle$, we obtain the short-time solutions of the coefficients
$\mathcal{A}$, $\mathcal{B}$, and $\mathcal{C}$ [see
Eq.~(\ref{A-ana})-Eq.(\ref{C-ana})]. Next, we focus on a time
regime: $\tau <S\tau ^{2}\ll1$ and $\gamma \tau \ll 1$, which allows
us to expand the above results in terms of $\beta$ ($\ll 1$)~\cite{Kitagawa}. From Eq.~(\ref{V+-}), it is easy to find that the product of the variances $V_{+}V_{-}=[(\mathcal{C}+\mathcal{A})(\mathcal{C}-\mathcal{A})-\mathcal{B}^2]/4$. To simplify it, we expand $\mathcal{C}\pm\mathcal{A}$, $\mathcal{B}^2$, and hence $V_{+}V_{-}$ up to the third-order of $\beta$ [see Eq.~(\ref{C+A})-Eq.~(\ref{V+V-ana})]. On the other hand, we can reduce the increased variance $V_{+}$ by keeping the lowest-order of $\beta$ (see Appendix B). Finally, using the relation $V_{-}=(V_{+}V_{-})/V_{+}$, we obtain analytical result of the reduced variance and also the
squeezing parameter:
\begin{equation}
\xi ^{2}\approx \frac{\gamma \tau }{\beta }+\frac{1}{4S\beta \sin
^{2}(\theta _{0})}+\frac{2\beta ^{2}}{3}\left[ 1+9S\sin ^{2}(\theta
_{0})\cos ^{2}(\theta _{0})\right] ,  \label{xi2ana}
\end{equation}
where $\beta$ is given by Eq.~(\ref{beta}) and $\theta _{0}$ is
polar angle of the initial state. In Fig.~\ref{fig1}, we compare
our analytical result of $\xi^{2}$ (red dash) with its exact
solution (solid line) for different values of the dephasing rate
$\gamma$. When $\gamma$ is not too large, we find that
Eq.~(\ref{xi2ana}) works well to predict the minimal value of the
squeezing parameter $\xi_{\min}^{2}=2V_{-}(\tau_{\min})/S$.
This is because both
the analytical and the exact results almost merge with each other
around the maximal-squeezing time $\tau_{\min}$.

Based upon Eq.~(\ref{beta}) and Eq.~(\ref{xi2ana}), we now analyze in detail the role of the phase dephasing on the spin
squeezing. If the first term on right-hand side of Eq.~(\ref{xi2ana}) is comparable with the second one, we obtain $\gamma\tau\sim[4S\sin
^{2}(\theta_{0})]^{-1}$. On the other hand, we compare the two terms of
Eq.~(\ref{beta}) and get $\gamma\sim S\tau\sin^{2}(\theta_{0})$.

\begin{figure} [htbp]
\centerline{\epsfig{file=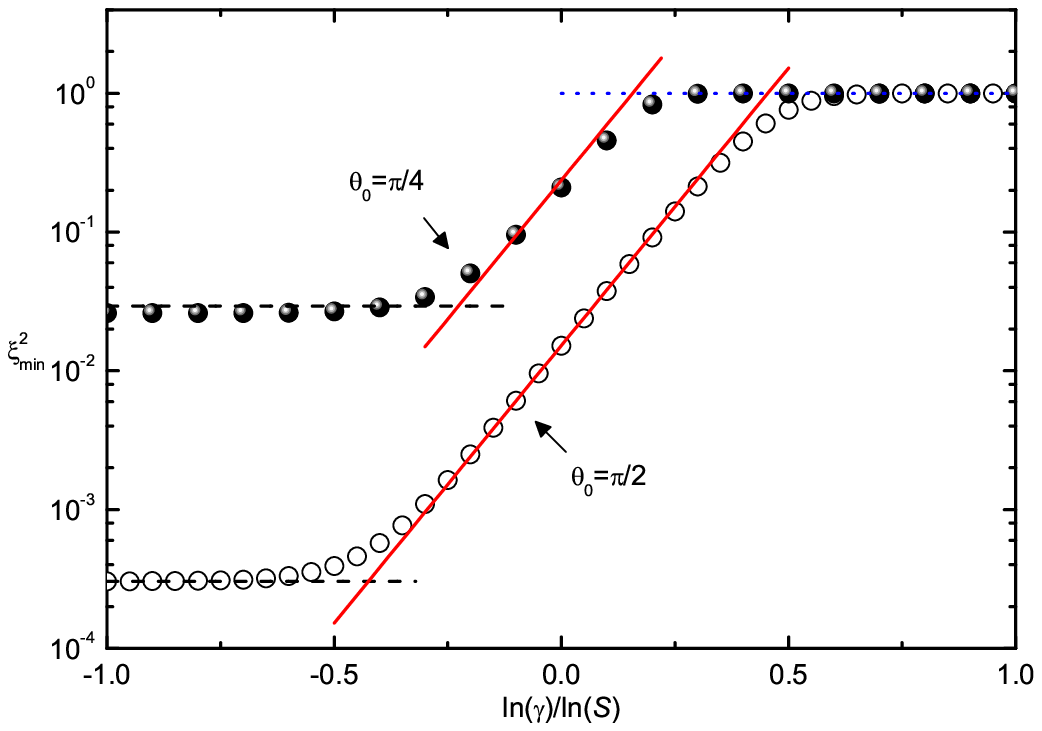, width=10.5cm}} 
\vspace*{13pt} \fcaption{\label{fig2}(color online) The minimal value of the squeezing
parameter $\xi _{\min }^{2}$ as a function of
$\ln(\protect\gamma)/\ln( S)$ for $\theta _{0}=\pi /2$ (open
circles) and $\theta _{0}=\pi /4$ (balls). The dash and the red
solid lines are obtained from our analytical results,
Eq.~(\ref{xi2min1}) and Eq.~(\ref{xi2min2}). The blue dotted line
corresponds to the case $\xi^2\approx1$ (almost no squeezing) as
$\ln(\protect\gamma)/\ln( S)>0.5$ for $\theta _{0}=\pi /2$ and
$\ln(\gamma)/\ln( S)>0.25$ for $\theta _{0}=\pi /4$. Other
parameters: $\chi =1$, $S=N/2=10^{5}$ and $\phi _{0}=0 $.}
\end{figure}

Obviously, the dephasing effect can be \textit{neglected} only if
$\gamma \ll S\tau\sin^{2}(\theta_{0})$ and $\gamma\tau\ll
[4S\sin^{2}(\theta_{0})]^{-1}$, for which Eq.~(\ref{beta}) becomes
$\beta\approx S\tau^{2}\sin^{2}(\theta _{0})$ and the first term
of Eq.~(\ref{xi2ana}) can be omitted. As a result, we obtain the
analytical result of the squeezing parameter~\cite{Jin09}:
\begin{equation}
\xi ^{2}\approx \frac{1}{4S\beta \sin ^{2}(\theta
_{0})}+\frac{2\beta^{2}}{3}\left[1+9S\sin^{2}(\theta _{0})\cos
^{2}(\theta _{0})\right] . \label{xi2ana1}
\end{equation}
From the relation $\left.(d\xi^{2}/d\tau)\right\vert_{\tau _{\min}}=0$,
we obtain the maximal-squeezing time:
\begin{equation}
\tau_{\min}=\chi t_{\min}\approx \frac{3^{1/6}[2S\sin ^{2}(\theta_{0})]^{-2/3}}{[1+9S\sin^{2}(\theta_{0})\cos^{2}(\theta_{0})]^{1/6}}.
\label{tmin1}
\end{equation}
Substituting Eq.~(\ref{tmin1}) back to Eq.~(\ref{xi2ana1}), we further
obtain the smallest value of $\xi^{2}$:
\begin{equation}
\xi_{\min}^{2}\approx\frac{3}{4}\left\{\frac{2\left[1+9S\sin^{2}(\theta_{0})\cos^{2}(\theta_{0})\right]}{3S^{2}\sin
^{4}(\theta_{0})}\right\}^{1/3}. \label{xi2min1}
\end{equation}
For the initial CSS with $\theta_{0}=\pi/2$, Eq.~(\ref{xi2min1}) shows $\xi_{\min}^{2}\approx\frac{1}{2}(\frac{2S}{3})^{-2/3}$, which is the best squeezing that the one-axis twisting scheme can reach~\cite{Kitagawa,Jin09}. Considering a large enough particle-number with $S=N/2=10^3$, we can obtain $\tau_{\min}\approx 7.565\times 10^{-3}$ and $\xi_{\min}^{2}\approx 6.552\times 10^{-3}$, fitting very well with numerical simulations [see Fig.~\ref{fig1}(a)]. From Eq.~(\ref{tmin1}), we find that the time scales as $\tau_{\min}\propto S^{-2/3}$ for $\theta
_{0}=\pi/2$~\cite{Kitagawa}, and $\tau_{\min}\propto S^{-5/6}$ for
$\theta_{0}\neq\pi/2$~\cite{Jin09}. Substituting the power rules into
the condition $\gamma\tau\ll[4S\sin^{2}(\theta_{0})]^{-1}$, we make a conclusion that our analytical results,
Eq.~(\ref{tmin1}) and Eq.~(\ref{xi2min1}), are valid for the
dephasing rate $\gamma\ll S^{-1/3}$ ($\theta_{0}=\pi/2$), or
$\gamma\ll S^{-1/6}$ ($\theta_{0}\neq\pi/2$). As shown by the
dash lines of Fig.~\ref{fig2}, one can find that
Eq.~(\ref{xi2min1}) coincides with the exact results when $\ln(\gamma)/\ln(S)<-0.5$.

To proceed, let us consider the case $\gamma<S\tau \sin
^{2}(\theta_{0})$, but $\gamma\tau>[4S\sin^{2}(\theta_{0})]^{-1}$, for which the first term of
Eq.~(\ref{xi2ana}) becomes important in a comparison with the
second one so we get
\begin{equation}
\xi ^{2}\approx \frac{\gamma\tau}{\beta}+\frac{2\beta^{2}}{3}\left[
1+9S\sin^{2}(\theta_{0})\cos^{2}(\theta_{0})\right] ,  \label{xi2ana2}
\end{equation}
where $\beta\approx S\tau^{2}\sin^{2}(\theta_{0})$. Minimizing $\xi^{2}$
with respect to $\tau$, we obtain
\begin{equation}
\tau_{\min}\approx\frac{(3\gamma)^{1/5}\left[8S^{3}\sin^{6}(\theta
_{0})\right]^{-1/5}}{[1+9S\sin^{2}(\theta_{0})\cos^{2}(\theta_{0})]^{1/5}},  \label{tmin2}
\end{equation}
and
\begin{equation}
\xi _{\min }^{2}\approx \frac{5}{4}\left\{ \frac{8\gamma
^{4}[1+9S\sin ^{2}(\theta _{0})\cos ^{2}(\theta _{0})]}{3S^{2}\sin
^{4}(\theta _{0})}\right\} ^{1/5}.  \label{xi2min2}
\end{equation}
From Eq.~(\ref{tmin2}), we find that the strongest squeezing
appears at $\tau\propto\gamma^{1/5}S^{-3/5}$ for $\theta
_{0}=\pi/2$, and $\tau\propto\gamma^{1/5}S^{-4/5}$ for $\theta
_{0}\neq\pi/2$. Using the conditions $\gamma\tau>[
4S\sin^{2}(\theta_{0})]^{-1}$ and $\gamma<S\tau\sin^{2}(\theta_{0})$, it is easy to find that Eq.~(\ref{xi2min2})
works quite well for a relatively weak dephasing rate with
$S^{-1/3}<\gamma<S^{1/2}$ ($\theta_{0}=\pi/2$), or
$S^{-1/6}<\gamma<S^{1/4}$ ($\theta_{0}\neq\pi/2$).

\begin{figure} [hpbt]
\centerline{\epsfig{file=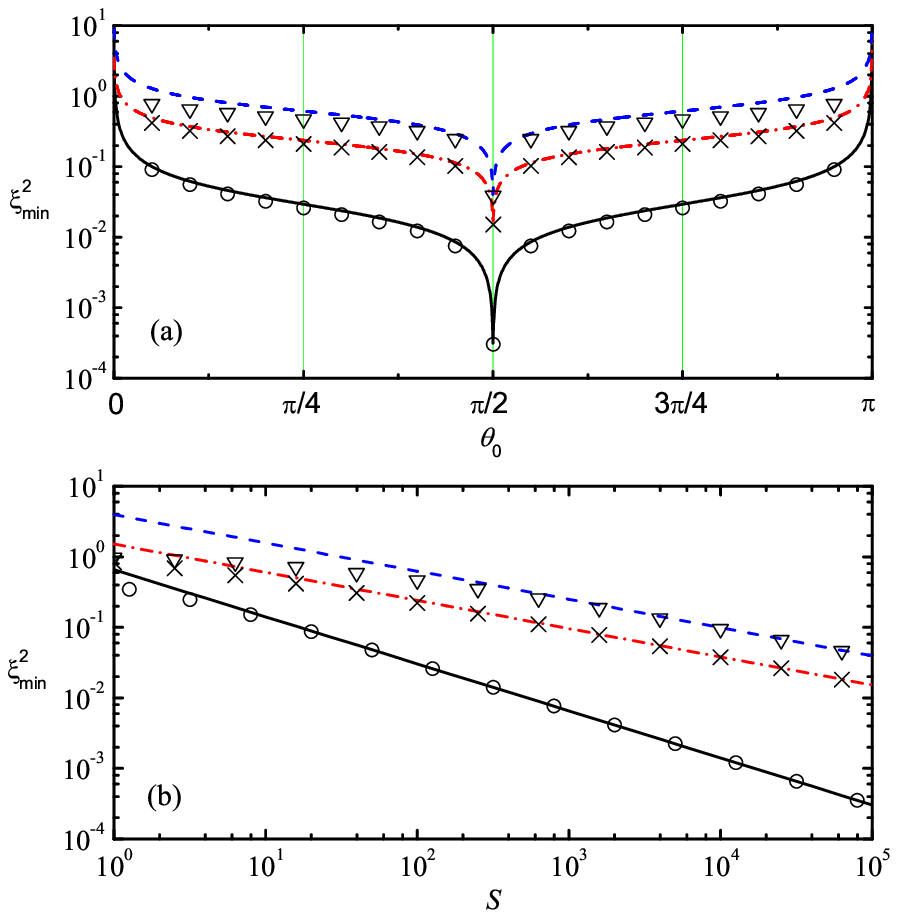, width=9.5cm}} 
\vspace*{13pt} \fcaption{\label{fig3}(Color online) The minimal value of the squeezing
parameter $\xi_{\min}^{2}$ against $\theta_{0}$
for $S=10^5$ (a), and $S$ for $\theta_{0}=\pi/2$ (b). From top to
bottom in each figure, the dephasing rates are taken as
$\gamma=3.3$ (triangles), $1$ (crosses), and $0$ (open circles).
The solid line is given by Eq.~(\ref{xi2min1}). The dashed and dashed-dot curves are predicted by
analytical result, Eq.~(\ref{xi2min2}). Other parameters are $\chi=1$ and $\phi_{0}=0$.}
\end{figure}

As shown in Fig.~\ref{fig1}(b), for the case $S=10^3$ and $\gamma=1$, Eq.~(\ref{tmin2})
and Eq.~(\ref{xi2min2}) predict $\tau_{\min}\approx 1.303\times 10^{-2}$ and $\xi_{\min}^{2}\approx 9.596\times 10^{-2}$, respectively. Both of
them fit with the exact numerical results $\tau_{\min}=1.264
\times 10^{-2}$ and $\xi_{\min}^{2}=9.278\times 10^{-2}$. As the dephasing rate increases up to $10$, our analytical results give $\tau_{\min}\approx 2.064\times 10^{-2}$ and $\xi_{\min}^{2}\approx 0.605$, which are in order-of-magnitude agreement with the exact results [see Fig.~\ref{fig1}(c)]. Moreover, we find that the red solid line of Fig.~\ref{fig2}, given by Eq.~(\ref{xi2min2})
for $\theta_{0}=\pi/4$, shows clearly that $\xi_{\min}^{2}$
increases rapidly with $\gamma$ in the regime $-0.2<\ln(\gamma)/\ln(S)<0.25$.

Our analytical results, Eq.~(\ref{xi2min1}) and Eq.~(\ref{xi2min2}), can not work to predict the spin
squeezing when $\gamma$ becomes large. For instance, let us
consider $\gamma\tau>[4S\sin^{2}(\theta_{0})]^{-1}$
and $\gamma>S\tau\sin^{2}(\theta_{0})$. In this case,
Eq.~(\ref{xi2ana2}) reduces to
\begin{equation}
\xi^{2}\approx 1+\frac{2\beta ^{2}}{3}\left[ 1+9S\sin ^{2}(\theta _{0})\cos
^{2}(\theta _{0})\right],  \label{xi2ana3}
\end{equation}
with $\beta\approx\gamma\tau$. Actually, when $\gamma>S^{1/2}$ ($\theta_{0}=\pi/2$), or $\gamma>S^{1/4}$
($\theta_{0}\neq\pi/2$), the squeezing effect is week due to $\xi
^{2}\approx 1$, as shown by the blue dotted line of Fig.~\ref{fig2}.

For a fixed $S$ ($=N/2$), we find from Fig.~\ref{fig3}(a) that the
achievable squeezing depends upon the polar angle of the
initial state; the initial CSS with $\theta_{0}=\pi/2$ is the
optimal one to minimize $\xi^{2}$ even in the presence of phase
dephasing. In Fig.~\ref{fig3}(b), we focus on the optimal case
$\theta_{0}=\pi/2$ and investigate the dependence of $\xi_{\min}^{2}$ on $S$ for the dephasing rates $\gamma=0$ (open circles),
$1$ (red crosses), and $3.3$ (blue triangles). The three curves, given by our analytical results, show good agreement with the exact numerical simulations for large enough particle number (say $S=N/2>10^{2}$). More interestingly, it is also found that
the slop of the solid curve is different with that of other two
lines. This is because our analytical
result for the case $\gamma =0$, Eq.~(\ref{xi2min1}), predicts $\xi_{\min}^{2}\propto
S^{-2/3}$~\cite{Kitagawa}; while for a small but nonzero $\gamma$, Eq.~(\ref{xi2min2}) gives $\xi_{\min}^{2}\propto S^{-2/5}$. Such a power rule has been also obtained by Takeuchi {\it et al.}~\cite{Takeuchi}. However, their scheme is based upon a double-pass Faraday interaction between atoms and far-off-resonant light. In addition, the starting point of their work, though quite similar with Eq.~(\ref{rhodp}), is derived by an approximated  commutation relation of the light-field Stokes operators~\cite{Takeuchi}.

\section{Conclusion}

In summary, we have investigated the role of phase diffusion on spin squeezing of the one-axis twisting model. Our results show that the spin squeezing depends upon the initial state $|\theta_{0}, 0\rangle$ ($=e^{-i\theta_0\hat{S}_y}|S, S\rangle$). The optimal initial state corresponds to the polar angle $\theta_0=\pi/2$, even in the presence of phase dephasing. If the dephasing rate $\gamma=\Gamma_p/\chi\ll S^{-1/3}$, the dephasing effect is negligible and the ideal one-axis twisting is restored. The strongest squeezing scales as $\xi^2_{\min}\propto S^{-2/3}$. For a moderate dephasing rate (i.e., $S^{-1/3}<\gamma<S^{1/2}$), the achievable squeezing obeys the power rule $\xi_{\min}^{2}\propto S^{-2/5}$, which is slightly worse than the ideal case. When the dephasing rate $\gamma>S^{1/2}$, we show that the squeezing becomes very weak due to $\xi^2\sim 1$.

\nonumsection{Acknowledgements} \noindent
We thank Professor L. You for helpful
discussions. This work is supported by Natural Science Foundation of China (NSFC, Contract No.~10804007 and No.~11174028), the Fundamental Research Funds for the Central Universities (Contract No.~2011JBZ013), and the Program for New Century Excellent Talents in University (Contract No.~NCET-11-0564). C.G.J is partially supported by National Innovation Experiment Program for
University Students (BJTU No.~1270021 and No.~1270037).

\appendix*

\section*{Exact solutions of the squeezing parameter}

As Eq.~(\ref{V+-}), the reduced and the increased variances depend upon the
coefficients $\mathcal{A}$, $\mathcal{B}$, and $\mathcal{C}$ (for details, see Ref.~\cite{Jin09}):
\begin{eqnarray}
\mathcal{A}&=&\frac{1}{2}\left\{\sin^{2}(\theta)\left[
S(S+1)-3\langle \hat{S}_{z}^{2}\rangle\right]-[1+\cos
^{2}(\theta )]\Re\left[\langle \hat{S}_{+}^{2}\rangle
e^{-2i\phi }\right] \right.  \nonumber \\
&&\left. +\sin (2\theta )\Re\left[\langle \hat{S}_{+}(2\hat{S}_{z}+1)\rangle e^{-i\phi }\right] \right\} ,  \label{A} \\
\mathcal{B}&=&-\cos (\theta)\Im\left[\langle
\hat{S}_{+}^{2}\rangle e^{-2i\phi }\right] +\sin (\theta
)\Im\left[\langle
\hat{S}_{+}(2\hat{S}_{z}+1)\rangle e^{-i\phi }\right],\label{B} \\
\mathcal{C}&=&S(S+1)-\langle \hat{S}_{z}^{2}\rangle
-\Re\left[\langle \hat{S}_{+}^{2}\rangle
e^{-2i\phi }\right] -\mathcal{A},  \label{C}
\end{eqnarray}
where the angles $\theta $ and $\phi $ are determined by the mean
spin $\langle\mathbf{S}\rangle=(\langle S_{x}\rangle, \langle
S_{y}\rangle, \langle S_{z}\rangle)$, with
\begin{eqnarray}
\sin (\theta)\!&=&\!\frac{|\langle \hat{S}_{+}\rangle |}{|\langle \mathbf{\hat{S}}\rangle|},
\hskip 8pt\cos (\theta )=\frac{\langle \hat{S}_{z}\rangle}{|\langle \mathbf{\hat{S}}\rangle |},  \label{theta} \\
\cos (\phi) \!&=&\! \frac{\langle \hat{S}_{x}\rangle }{|\langle
\hat{S}_{+}\rangle|},
\hskip 8pt\sin (\phi)=\frac{\langle
\hat{S}_{y}\rangle}{|\langle\hat{S}_{+}\rangle|}.
\label{sinphi}
\end{eqnarray}
Here, $\langle\hat{S}_{x}\rangle=\Re\langle\hat{S}_{+}\rangle$ and $\langle\hat{S}_{y}\rangle=\Im
\langle\hat{S}_{+}\rangle$, as mentioned above.
Note that the above formulae are valid for any SU(2) system.
Moreover, one can find that $\mathcal{A}$, $\mathcal{B}$, and
$\mathcal{C}$ depend only on five expectation values: $\langle
\hat{S}_{z}\rangle$, $\langle\hat{S}_{+}\rangle$, $\langle
\hat{S}_{z}^{2}\rangle$, $\langle \hat{S}_{+}^{2}\rangle$, and
$\langle \hat{S}_{+}(2\hat{S}_{z}+1)\rangle$. Although the
argument $\phi$ appears in the coefficients, we do \textit{not} need to know
its explicit expression. Instead, only $\cos(\phi)$ and $\sin(\phi)$ are needed in real calculations of the coefficients and hence
the variances $V_{\pm}$.

Starting with an initial CSS $|\theta_{0}, \phi_{0}\rangle$, the OAT model can be solved exactly. For instance, we calculate the expectation value $\langle\hat{S}_{+}^{l}\rangle\equiv\mathrm{Tr}(\hat{\rho}\hat{S}_{+}^{l})$, with an integer $l=1, 2, \cdots, [S]$. Here, $[S]$ denotes the greatest
integer of any real number $S$. Using Eq.~(\ref{rhodp}), we obtain
\begin{eqnarray}
\langle \hat{S}_{+}^{l}\rangle
=\sum_{m=-S}^{S-l}\rho
_{m,m+l}(0)X_{-m}X_{-m-1}\cdots X_{-m-l+1}e^{i(2ml+l^{2})\tau
}e^{-\gamma l^{2}\tau}=e^{-\gamma l^{2}\tau}\langle
\hat{S}_{+}^{l}\rangle _{0},  \label{Jpl}
\end{eqnarray}
where $X_{m}=\sqrt{(S+m)(S-m+1)}$, and $\rho
_{m,n}(0)=c_{m}c_{n}^{\ast}$ represent the
density-matrix elements of the initial CSS, with $c_{m}$ given by Eq.~(\ref{cm}). In addition, we have introduced
\begin{eqnarray}
\langle \hat{S}_{+}^{l}\rangle _{0} &\equiv &\sum_{m=-S}^{S-l}\rho
_{m,m+l}(0)e^{i(2ml+l^{2})\tau }X_{-m}X_{-m-1}\cdots X_{-m-l+1}  \nonumber \\
&=&\sum_{m=-S}^{S-l}\frac{(2S)!}{(S+m)!(S-m-l)!}e^{il\phi
_{0}}e^{i(2ml+l^{2})\tau }\left( \cos \frac{\theta _{0}}{2}\right)
^{2S+2m+l}\left( \sin \frac{\theta _{0}}{2}\right) ^{2S-2m-l} \nonumber \\
&=&\frac{(2S)!}{2^{l}(2S-l)!}\sin ^{l}\left(\theta _{0}\right)e^{il\phi
_{0}}
\sum_{m}{2S-l\choose S+m} \left(e^{il\tau }\cos ^{2}\frac{\theta _{0}}{2}
\right)^{S+m}\left(e^{-il\tau}\sin ^{2}\frac{\theta
_{0}}{2}\right)^{S-m-l}  \nonumber \\
&=&\frac{(2S)!}{2^{l}(2S-l)!}\sin ^{l}(\theta _{0})e^{il\phi
_{0}}[R(l\tau )]^{2S-l},
\label{Jpl0}
\end{eqnarray}
where $R(l\tau)=\cos(l\tau)+i\cos(\theta_{0})\sin(l\tau)$, and we have used the binomial formula:
\begin{equation}
\sum_{m=-S}^{S-l}{2S-l\choose S+m}a^{S+m}b^{S-m-l}=\sum_{n=0}^{2S-l}{2S-l\choose n}a^{n}b^{2S-n-l}=(a+b)^{2S-l}.
\end{equation}
With the help of Eq.~(\ref{Jpl}) and
Eq.~(\ref{Jpl0}), we can obtain the exact solutions of $\langle\hat{S}_{+}\rangle$ and
$\langle\hat{S}_{+}^{2}\rangle$, given by Eq.~(\ref{JzJ+}) and
Eq.~(\ref{S+2exact}), respectively.

We note that without the dephasing, the spin system evolves under governed by the OAT Hamiltonian $\hat{H}=\chi\hat{S}_z^2$, so we have $d\langle\hat{S}_{+}\rangle_0/d\tau=-i\langle[\hat{S}_{+}, \hat{H}]\rangle_0/\chi=i\langle\hat{S}_{+}(2\hat{S}_z+1)\rangle_0$, where the subscript $0$ denotes the expectation values in the absence of phase dephasing (i.e., $\gamma=0$), and $\langle\hat{S}_{+}\rangle_{0}$ has been given in Eq.~(\ref{Jpl0}) with $l=1$. Therefore, we obtain
\begin{eqnarray}
\langle\hat{S}_{+}(2\hat{S}_{z}+1)\rangle=e^{-\gamma \tau}\langle\hat{S}_{+}(2\hat{S}_{z}+1)\rangle_{0}=-ie^{-\gamma\tau}\frac{d\langle \hat{S}_{+}\rangle_{0}}{d\tau},
\end{eqnarray}
which gives Eq.~(\ref{S+2Jz+1exact}). Finally, we calculate the population imbalance and its variance:
\begin{eqnarray}
\langle\hat{S}_{z}\rangle &=&\sum_{m=-S+1}^{S}(S+m)\rho
_{m,m}(0)-\sum_{m=-S}^{S}S\rho _{m,m}(0)  \nonumber \\
&=&2S\cos ^{2}\left(\frac{\theta _{0}}{2}\right)
\sum_{m}{2S-1 \choose S+m-1}\left(\cos ^{2}
\frac{\theta _{0}}{2}\right)^{S+m-1}\left( \sin
^{2}\frac{\theta _{0}}{2}\right)^{S-m}-S  \nonumber\\
&=&2S\cos ^{2}\left( \frac{\theta _{0}}{2}\right) -S=S\cos (\theta _{0}),
\label{Jz}
\end{eqnarray}
and
\begin{eqnarray}
\langle \hat{S}_{z}^{2}\rangle &=&\sum_{m=-S}^{S}S^{2}\rho
_{m, m}(0)-\sum_{m=-S+1}^{S-1}(S+m)(S-m)\rho _{m,m}(0)  \nonumber \\
&=&S^{2}-S\left(S-\frac{1}{2}\right)\sin^{2}(\theta _{0}) \sum_{m}{2S-2\choose S+m-1}\left(\cos ^{2}
\frac{\theta _{0}}{2}\right)^{S+m-1}\left( \sin
^{2}\frac{\theta_{0}}{2}\right)^{S-m-1}  \nonumber \\
&=&S^{2}-S\left( S-\frac{1}{2}\right) \sin ^{2}\left( \theta
_{0}\right) =\frac{S}{2}+S\left( S-\frac{1}{2}\right) \cos
^{2}(\theta _{0}),  \label{Jz2}
\end{eqnarray}
where we have used the normalization condition $\sum_{m}\rho_{m,m}(0)=1$. So far we have solved all the quantities that relevant to get the coefficients $\mathcal{A}$, $\mathcal{B}$, and
$\mathcal{C}$, with which we can calculate exactly the variances $V_{\pm}$, and hence the squeezing parameter $\xi^{2}$.

\appendix*
\section*{Short-time solutions of the squeezing parameter}

To obtain analytical results of the coefficients $\mathcal{A}$,
$\mathcal{B}$, and $\mathcal{C}$, we have to make further
approximations~\cite{Jin09}, $\sin(\theta )=|\langle
\hat{S}_{+}\rangle|/|\langle\mathbf{\hat{S}}\rangle|\approx
\sin(\theta_{0})$ and $\cos(\theta)=\langle\hat{S}_{z}\rangle
/|\langle\mathbf{\hat{S}}\rangle|\approx\cos(\theta_{0})$,
where $\theta_{0}$ is polar angle of the initial CSS.
Substituting Eq.~(\ref{S+2a}) and Eq.(\ref{S+2Jz+1a}), as well as Eq.(\ref{Jz2}) into Eq.~(\ref{A}), we obtain the short-time solution of the coefficient
\begin{eqnarray}
\mathcal{A} &\approx &\frac{\sin ^{2}\theta _{0}}{2}\left\{ \left[
S(S+1)-3\left( \frac{S}{2}+S\left( S-\frac{1}{2}\right) \cos ^{2}\theta
_{0}\right) \right] \right.   \nonumber \\
&&\left. -S\left( S-\frac{1}{2}\right) \left( 1+\cos ^{2}\theta _{0}\right)
e^{-4\beta }+4S\left( S-\frac{1}{2}\right) e^{-\beta }\cos ^{2}\theta
_{0}\right\}   \nonumber \\
&=&S\left( S-\frac{1}{2}\right) \frac{\sin ^{2}\theta _{0}}{2}\left\{
1-e^{-4\beta }-\left( 3+e^{-4\beta }-4e^{-\beta }\right) \cos ^{2}\theta
_{0}\right\} .  \label{A-ana}
\end{eqnarray}%
Hereafter, we assume $S(S-1/2)\approx S^{2}$ for large enough particle number $N$ ($=2S>100$). Similarly,
we have $\mathcal{B}\approx 2S^{2}\tau\sin^{2}(\theta_{0})e^{-\beta}$,
and
\begin{equation}
\mathcal{C}\approx \frac{S^{2}\sin^{2}\theta_{0}}{2}\left\{1-e^{-4\beta
}+(3+e^{-4\beta}-4e^{-\beta})\cos^{2}\theta_{0}\right\}+S.
\label{C-ana}
\end{equation}

Following Kitagawa and Ueda~\cite{Kitagawa}, we focus on a time
regime $\tau <S\tau^{2}\ll1$ and $\gamma \tau\ll1$, which allows
us to expand the coefficients $\mathcal{A}$, $\mathcal{B}$,
$\mathcal{C}$ in terms of $\beta $ because of $\beta\ll1$.
Firstly, we calculate the product of the variances $V_{+}V_{-}$,
which is given by Eq.~(\ref{V+-}),
\begin{equation}
V_{+}V_{-}=\frac{1}{4}\left[(\mathcal{C}+\mathcal{A})(\mathcal{C}-\mathcal{A})-\mathcal{B}^{2}\right].
\label{V+V-a}
\end{equation}
To calculate it, we expand $\mathcal{C}\pm \mathcal{A}$ and
$\mathcal{B}^{2}$ up to the third-order of $\beta $ (also $\gamma
\tau $) and obtain
\begin{eqnarray}
&&\mathcal{B}^{2}\approx 4S^{3}\sin ^{2}(\theta_{0})\left(\beta-\gamma\tau\right)\left(1-2\beta +2\beta ^{2}\right) , \nonumber\\
&&\mathcal{C}+\mathcal{A} \approx 4S^{2}\sin^{2}(\theta_{0})\beta\left(1-2\beta+\frac{8}{3}\beta^{2}\right)+S, \label{C+A} \\
&&\mathcal{C}-\mathcal{A} \approx 6S^{2}\sin^{2}(\theta _{0})\cos^{2}(\theta_{0})\beta^2\left(1-\frac{5}{3}\beta\right) +S. \nonumber
\end{eqnarray}%
Keeping the terms up to $O[(S\beta)^{3}]$, we obtain
\begin{eqnarray}
V_{+}V_{-} &\approx &\frac{S^{2}}{4}\left\{1+4S\sin ^{2}(\theta _{0})\left[
\frac{2}{3}\beta ^{3} +6S\sin ^{2}(\theta _{0})\cos ^{2}(\theta _{0})\beta ^{3}\right.
\right.  \nonumber \\
&&\left. \left. + \gamma \tau+\frac{3}{2}\cos ^{2}(\theta
_{0})\beta ^{2}\left( 1-\frac{5}{3}\beta \right) \right] \right\} .
\label{V+V-ana}
\end{eqnarray}
For brevity, we will omit the last term $3\cos^{2}(\theta_{0})\beta^{2}(\cdots)/2$, though its contribution may be larger
than that of the term $2\beta^{3}/3$.

Next, we calculate the increased variance $V_{+}$. From
Eq.~(\ref{A-ana})-Eq.~(\ref{C-ana}), we note that the leading
terms of the coefficients $\mathcal{A}\propto S^{2}\beta$ ($\propto
S^{3}\tau^{2}$) and $\mathcal{B}\propto S^{2}\tau$. In the time scale with $S^{2}\tau>1$, it is easy to find that
$\mathcal{A}>\mathcal{B}$, so the increased variance can
be simplified as
\begin{equation}
V_{+}\approx \frac{1}{2}(\mathcal{C}+\mathcal{A})\approx 2S^{2}\sin
^{2}(\theta _{0})\beta.  \label{V+ana}
\end{equation}
where we only keep the lowest-order of $\beta$ in the last step [see also Eq.~(\ref{C+A})]. Finally,
using $V_{-}=(V_{+}V_{-})/V_{+}$, we obtain analytical result of the reduced
variance
\begin{equation}
V_{-}\approx \frac{S}{2}\left\{ \frac{1}{4S\sin ^{2}(\theta
_{0})\beta }+\frac{\gamma \tau }{\beta }+\frac{2}{3}\beta
^{2}\left[ 1+\frac{9S}{4}\sin ^{2}(2\theta _{0})\right] \right\} ,
\label{V-ana}
\end{equation}
which gives the analytical result of the squeezing parameter, {\it i.e.}, Eq.~(\ref{xi2ana}).

\nonumsection{References}


\noindent


\begin{thebibliography}{000}

\bibitem{Caves} C. M. Caves (1981), \textit{Quantum-mechanical noise in an interferometer}, Phys. Rev. D, 23, pp. 1693-1708.

\bibitem{Yurke} B. Yurke, S. L. McCall, and S. R. Klauder (1986), \textit{SU(2) and SU(1,1) interferometers}, Phys. Rev. A, 33, pp. 4033-4054.

\bibitem{Wineland} D. J. Wineland, J. J. Bollinger, W. M. Itano, and F. L. Moore (1992), \textit{Spin squeezing and reduced quantum noise in spectroscopy}, Phys. Rev. A, 46, pp. R6797-R6800; D. J. Wineland, J. J. Bollinger, W. M. Itano, and F. L. Moore (1994), \textit{Squeezed atomic states and projection noise in spectroscopy}, Phys. Rev. A, 50, pp. 67-88.

\bibitem{Kitagawa} M. Kitagawa and M. Ueda (1993), \textit{Squeezed spin states}, Phys. Rev. A, 47, pp. 5138-5143.

\bibitem{Ma} J. Ma, X. Wang, C. P. Sun, and F. Nori (2011), \textit{Quantum spin squeezing}, Phys. Reports, 509, pp. 89-166.

\bibitem{Molmer} K. M{\o}lmer and A. S\o rensen (1999), \textit{Multiparticle Entanglement of Hot Trapped Ions}, Phys. Rev. Lett., 82, pp. 1835-1838.

\bibitem{You03a} L. You (2003), \textit{Creating Maximally Entangled Atomic States in a Bose-Einstein Condensate}, Phys. Rev. Lett., 90, pp. 030402.

\bibitem{Pezze&Smerzi} L. Pezz\'{e} and A. Smerzi (2009), \textit{Entanglement, Nonlinear Dynamics, and the Heisenberg Limit}, Phys. Rev. Lett., 102, pp. 100401.

\bibitem{YCL&YL} Y. C. Liu, Z. F. Xu, G. R. Jin, and L. You (2011), \textit{Spin Squeezing: Transforming One-Axis Twisting into Two-Axis Twisting}, Phys. Rev. Lett., 107, pp. 013601.
\bibitem{Milburn} G. J. Milburn, J. Corney, E. M. Wright, and D. F. Walls (1997), \textit{Quantum dynamics of an atomic Bose-Einstein condensate in a double-well potential}, Phys. Rev. A, 55, pp. 4318-4324.

\bibitem{Cirac} J. I. Cirac, M. Lewenstein, K. M{\o}lmer, and P. Zoller (1998), \textit{Quantum superposition states of Bose-Einstein condensates}, Phys. Rev. A, 57, pp. 1208-1218.

\bibitem{Sorensen} A. S\o rensen, L.-M. Duan, J. I. Cirac, and P. Zoller (2001), \textit{Many-particle entanglement with Bose¨CEinstein condensates}, Nature, 409, pp. 63-66; U. V. Poulsen and K. M{\o }lmer(2001), \textit{Positive-P simulations of spin squeezing in a two-component Bose condensate}, Phys. Rev. A, 64, pp. 013616.

\bibitem{Bigelow} S. Raghavan, H. Pu, P. Meystre, and N. P. Bigelow (2001), \textit{Generation of arbitrary Dicke states in spinor Bose-Einstein condensates}, Opt. Commu., 188, pp. 149-154.

\bibitem{Law} C. K. Law, H. T. Ng, and P. T. Leung (2001), \textit{Coherent control of spin squeezing}, Phys. Rev. A, 63, pp. 055601.

\bibitem{Jin07} G. R Jin and S. W. Kim (2007), \textit{Storage of Spin Squeezing in a Two-Component Bose-Einstein Condensate}, Phys. Rev. Lett., 99, pp. 170405; G. R Jin and S. W. Kim (2007), \textit{Spin squeezing and maximal-squeezing time}, Phys. Rev. A, 76, pp. 043621.

\bibitem{Jin08} G. R Jin and C. K. Law (2008), \textit{Relationship between spin squeezing and single-particle coherence in two-component Bose-Einstein condensates with Josephson coupling}, Phys. Rev. A, 78, pp. 063620; G. R. Jin, X. W. Wang, D. Li, and Y. W. Lu (2010), \textit{Atom-number squeezing and bipartite entanglement of two-component Bose-Einstein condensates: analytical results}, J. Phys. B, 43, pp. 045301.

\bibitem{YC} J. Grond, J. Schmiedmayer, and U. Hohenester (2009), \textit{Optimizing number squeezing when splitting a mesoscopic condensate}, Phys. Rev. A, 79, pp. 021603(R).

\bibitem{Orzel} P. Bouyer and M. A. Kasevich (1997), \textit{Heisenberg-limited spectroscopy with degenerate Bose-Einstein gases}, Phys. Rev. A, 56, pp. R1083-R1086; C. Orzel, A. K. Tuchman, M. L. Fenselau, M. Yasuda, M. A. Kasevich (2001), Science, 291, pp. 2386-2389; W. Li, A. K. Tuchman, H. C. Chien, and M. A. Kasevich (2007), \textit{Extended Coherence Time with Atom-Number Squeezed States}, Phys. Rev. Lett., 98, pp. 040402.

\bibitem{Greiner} M. Greiner, O. Mandel, T. Esslinger, T. W. H\"{a}nsch, and I. Bloch (2002), \textit{Quantum phase transition from a superfluid to a Mott insulator in a gas of ultracold atoms}, Nature, 415, pp. 39-44.

\bibitem{Strabley} J. Sebby-Strabley, B. L. Brown, M. Anderlini, P. J. Lee, W. D. Phillips, and J. V. Porto (2007), \textit{Preparing and Probing Atomic Number States with an Atom Interferometer}, Phys. Rev. Lett., 98, pp. 200405.

\bibitem{Jaksch} M. Rodr\'{\i}guez, S. R. Clark, and D. Jaksch (2007), \textit{Generation of twin Fock states via transition from a two-component Mott insulator to a superfluid}, Phys. Rev. A, 75, pp. 011601(R).

\bibitem{Esteve} J. Esteve, C. Gross, A. Weller, S. Giovanazzi, and M. K. Oberthaler (2008), \textit{Squeezing and entanglement in a Bose¨CEinstein condensate}, Nature, 455, pp. 1216-1219 .

\bibitem{Chuu} C.-S. Chuu, F. Schreck, T. P. Meyrath, J. L. Hanssen, G. N. Price, and M. G. Raizen (2005), \textit{Direct Observation of Sub-Poissonian Number Statistics in a Degenerate Bose Gas}, Phys. Rev. Lett., 95, pp. 260403.

\bibitem{Jo} G.-B. Jo, Y. Shin, S. Will, T. A. Pasquini, M. Saba, W. Ketterle, and D. E. Pritchard (2007), \textit{Long Phase Coherence Time and Number Squeezing of Two Bose-Einstein Condensates on an Atom Chip}, Phys. Rev. Lett., 98, pp. 030407.

\bibitem{Gross} C. Gross, T. Zibold, E. Nicklas, J. Est\`{e}ve, and M. K. Oberthaler (2010), \textit{Nonlinear atom interferometer surpasses classical precision limit}, Nature, 464, pp. 1165-1169.

\bibitem{Riedel} M. F. Riedel, P. Bohi, Y. Li, T. W. Hansch, A. Sinatra, and P. Treutlein (2010),\textit{Atom-chip-based generation of entanglement for quantum metrology}, Nature, 464, pp. 1170-1173.

\bibitem{Maussang} K. Maussang, G. E. Marti, T. Schneider, P. Treutlein, Y. Li, A. Sinatra, R. Long, J. Est\`{e}ve, and J. Reiche (2010), \textit{Enhanced and Reduced Atom Number Fluctuations in a BEC Splitter}, Phys. Rev. Lett., 105, pp. 080403.

\bibitem{LiYun} Y. Li, P. Treutlein, J. Reichel, and A. Sinatra (2009), {\it Spin squeezing in a bimodal condensate: spatial dynamics and particle losses}, Eur. Phys. J. B., 68, pp. 365-381; Y. Li, Y. Castin, and A. Sinatra (2008), {\it Optimum Spin Squeezing in Bose-Einstein Condensates with Particle Losses}, Phys. Rev. Lett., 100, pp. 210401.

\bibitem{Jin09} G. R. Jin, Y. C. Liu, and W. M. Liu (2009), \textit{Spin squeezing in a generalized one-axis twisting model}, New J. Phys., 11, pp. 073049.

\bibitem{Gardiner} C. W. Gardiner and P. Zoller (2000), \textit{Quantum Noise}, Springer (Berlin).

\bibitem{Milburn08} D. F. Walls and G. J. Milburn (2008), \textit{Quantum Optics}, Springer-Verlag (Berlin
Heidelberg).

\bibitem{Huelga} S. F. Huelga, C. Macchiavello, T. Pellizzari, A. K. Ekert, M. B. Plenio, and J. I. Cirac (1997), \textit{Improvement of Frequency Standards with Quantum Entanglement}, Phys. Rev. Lett., 79, pp. 3865-3868.

\bibitem{Kitagawa01} D. Ulam-Orgikh and M. Kitagawa (2001), \textit{Spin squeezing and decoherence limit in Ramsey spectroscopy}, Phys. Rev. A, 64, pp. 052106.

\bibitem{Rey} A. M. Rey, L. Jiang, and M. D. Lukin (2007), \textit{Quantum-limited measurements of atomic scattering properties}, Phys. Rev. A, 76, pp. 053617.

\bibitem{Boixo08} S. Boixo, A. Datta, S. T. Flammia, A. Shaji, E. Bagan, and C. M. Caves (2008), \textit{Quantum-limited metrology with product states}, Phys. Rev. A, 77, pp. 012317.

\bibitem{Khodorkovsky} Y. Khodorkovsky, G. Kurizki, and A. Vardi (2009), \textit{Decoherence and entanglement in a bosonic Josephson junction: Bose-enhanced quantum Zeno control of phase diffusion}, Phys. Rev. A, 80, pp. 023609.

\bibitem{ljy} Y. C. Liu, G. R. Jin, and L. You (2010), \textit{Quantum-limited metrology in the presence of collisional dephasing}, Phys. Rev. A, 82, pp. 045601.

\bibitem{Jin10} G. R. Jin, Y. C. Liu, and L. You (2011), \textit{Optimal phase sensitivity of atomic Ramsey interferometers with coherent spin states}, Front. of Phys., 6, pp. 251-257.

\bibitem{Artur} A. Widera, S. Trotzky, P. Cheinet, S. F\"{o}lling, F. Gerbier, I. Bloch, V. Gritsev, M. D. Lukin, and E. Demler (2008),
\textit{Quantum Spin Dynamics of Mode-Squeezed Luttinger Liquids in Two-Component Atomic Gases}, Phys. Rev. Lett., 100, pp. 140401.




\bibitem{Agarwal} R. R. Puri and G. S. Agarwal (1992), \textit{Exact density matrix for the degenerate-Raman-coupled model in the presence of collisions}, Phys. Rev. A, 45, pp. 5073-5077.

\bibitem{Chen} T. W. Chen and P. T. Leung (2003), \textit{Effect of collision dephasing on atomic evolutions in a high-Q cavity}, Phys. Rev. A, 67, pp. 055802.

\bibitem{CSS} S. M. Radcliffe (1971), \textit{Some properties of coherent spin states}, S. Phys. A, 4, pp. 313; F. T. Arecchi, E. Courtens, R. Gilmore, and H. Thomas (1972), \textit{Atomic Coherent States in Quantum Optics}, Phys. Rev. A, 6, pp. 2211-2237; W.-M. Zhang, D. H. Feng, and R. Gilmore (1990), \textit{Coherent States: Theory and Some Applications}, Rev. Mod. Phys., 62, pp. 867-928.


\bibitem{Takeuchi} M. Takeuchi, S. Ichihara1, T. Takano1, M. Kumakura, T. Yabuzaki, and Y. Takahashi (2005), \textit{Spin Squeezing via One-Axis Twisting with Coherent Light}, Phys. Rev. Lett., 94, pp. 023003.

\bibitem{Genoni} M. G. Genoni, S. Olivares, and M. G. A. Paris (2011), \textit{Optical phase estimation in the presence of phase-diffusion}, (2011) Phys. Rev. Lett. 106, pp. 153603; M. G. Genoni, S. Olivares, D. Brivio, S. Cialdi, D. Cipriani, A. Santamato, S. Vezzoli, and M. G. A. Paris (2012), \textit{Optical interferometry in the presence of large phase diffusion}, Phys. Rev. A., 85, pp. 043817.

\bibitem{Ferrini} G. Ferrini, D. Spehner, A. Minguzzi, and F. W. J. Hekking (2011), \textit{Effect of phase noise on useful quantum correlations in Bose Josephson junctions}, Phys. Rev. A., 84, pp. 043628.

\bibitem{Sinatra1} A. Sinatra, J.-C. Dornstetter, and Y. Castin (2012), \textit{Spin squeezing in Bose-Einstein condensates: Limits imposed by decoherence and non-zero temperature}, Front. Phys., 7, pp. 86-97.


\bibitem{Satio} M. Ueda, T. Wakabayashi, and M. Kuwata-Gonokami (1996), \textit{Synchronous Collapses and Revivals of Atomic Dipole Fluctuations and Photon Fano Factor beyond the Standard Quantum Limit}, Phys. Rev. Lett., 76, pp. 2045-2048; H. Saito and M. Ueda (1999), \textit{Squeezed few-photon states of the field generated from squeezed atoms}, Phys. Rev. A, 59, pp. 3959-3974; H. Saito and M. Ueda (1997), \textit{Quantum-Controlled Few-Photon State Generated by Squeezed Atoms}, Phys. Rev. Lett., 79, pp. 3869-3872.

\bibitem{Sinatra} A. Sinatra and Y. Castin (2000), \textit{Binary mixtures of Bose-Einstein condensates: Phase dynamics and spatial dynamics}, Eur. Phys. J. D, 8, pp. 319-332.


\end{thebibliography}
\end{document}